\documentclass{aastex}
\usepackage{spr-astr-addons}
\usepackage{url}
\RequirePackage{color}

\shorttitle{Global Spectral Energy Distributions of the Large Magellanic Cloud with Interstellar Dust}
\shortauthors{Kim et al.}

\slugcomment{Accepted to Astrophysics and Space Science}
\begin{document}

\title{Global Spectral Energy Distributions of the Large Magellanic Cloud with Interstellar Dust}

\author{Sungeun Kim\altaffilmark{1}} \and \author{Eunjoo Kwon\altaffilmark{1,*}} \and \author{Kyoung-Sook Jeong\altaffilmark{1}} \and \author{Kihun Kim\altaffilmark{1}} \and \author{Chiyoung Cho\altaffilmark{1}} \and \author{Eun Jung Chung\altaffilmark{1}}

\begin{abstract}

The effects of dust on infrared emission vary among galaxies of different morphological types. We investigated integrated spectral energy distributions (SEDs) in infrared and submillimeter/millimeter emissions from the Large Magellanic Cloud (LMC) based on observations from the Herschel Space Observatory (HSO) and near- to mid-infrared observations from the Spitzer Space Telescope (SST). We also used IRAS and WMAP observations to constrain the SEDs and present the results of radiative transfer calculations using the spectrophotometric galaxy model. We explain the observations by using dust models with different grain size distributions in the interstellar medium of the LMC, noting that the LMC has undergone processes that differ from those in the Milky Way. We determined a spectral index and a normalisation factor in the range of $-3.5$ to $-3.45$ with grain radii in the range of 1 nm -- 300 nm for the silicate grain and 2 nm -- 1 $\mu$m for the graphite grain. The best fit to the observed SED was obtained with a spectral index of $-3.47$. The grain size distribution is described using a power law but with a break that is introduced below $a_b$, where a larger exponent is used. Changing the graphite grain size distribution significantly changed the SED pattern within the observational uncertainties. Based on the SED fits to the observations from submillimeter wavelengths to infrared radiation from the LMC using GRASIL (Silva et al. 1998), we obtained a reasonable set of parameter values in chemical and geometric space together with the grain size distributions (Weingartner \& Draine 2001) and a modified MRN model with the LMC extinction curve (Piovan et al. 2006a). For a given set of parameters including the disc scale height, synthesis of the starlight spectrum, optical depth, escape time scale, dust model, and star formation efficiency, the adopted dust-to-gas ratio for modeling the observed SEDs, $\sim1/300$ (from the literature) yields a reasonable fit to the observed SEDs and similar results with the metallicity of the LMC as those reported in Russell \& Bessell (1989). The dust-to-gas ratios that are given as the metallicity caused the variation in the model fits. The difference mainly appears at the wavelength near 100 $\mu$m.
\end{abstract}
\keywords{Interstellar medium: dust; Stars: star formation; Dust: grain size distribution; Galaxies: spectral energy distributions; Galaxy: Large Magellanic Cloud}
\altaffiltext{1}{Department of Astronomy and Space Science, Sejong University, 98 Gunja-dong, Gwangjin-gu, 143-747, Seoul, South Korea; E-mail: sek@sejong.ac.kr}
\altaffiltext{*}{Current Address: Korea Aerospace Research Institute, Satellite Information Research Center, 169-84, Gwahak-ro, Yuseong-gu, Daejeon 305-806, South Korea}

\section{Introduction}

Dust grains in the interstellar medium (ISM) of galaxies absorb and scatter stellar radiation, mainly at ultraviolet (UV) and optical wavelengths, and re-emit in the far-infrared. Infrared observations from satellite telescopes have revealed that dust grains play an important role in reprocessing a significant amount of stellar radiation in the local universe. An appropriate treatment of dust reprocessing in galaxies is essential to determine physical quantities, including the star formation rates (SFRs) and star formation histories from observed data, as well as to test theoretical galaxy formation models against these observations.

To date, information about the composition, morphology, size distribution, and relative abundance of various constituents of interstellar dust has been provided by observations of our own Galaxy, the Large Magellanic Cloud (LMC), and nearby galaxies. Various models have been proposed to explain the properties of interstellar dust (Mathis, Rumpl, \& Nordsiek 1977; Draine \& Lee 1984; D\`{e}sert et al. 1990; Dopita et al. 1995; Draine \& Li 2001; Weingartner \& Draine 2001; Zubko, Dwek, \& Arendt 2004; Draine \& Li 2007). Based on near-infrared to far-infrared observations, the effects of dust on infrared emission appear to depend on chemical and physical composition, star formation rate, dust-to-gas mass ratio, metallicity, and star formation efficiency (Silva et al. 1998; Piovan et al. 2006a,b), as well as the effects of grain heating by starlight (Draine \& Li 2007). The size distribution of dust grains has an important effect on the spectral energy distribution (SED) models for the diffuse interstellar medium (ISM) (Mathis, Rumpl, \& Nordsiek 1977; Draine \& Lee 1984; Kim et al. 1994; Li \& Draine 2001). The effects of dust on the infrared emission can also vary between galaxies of different morphological types.

In this paper, we report the integrated SEDs from the infrared and submillimeter/millimeter emissions in the LMC and describe the results of radiative transfer calculations using the spectrophotometric galaxy model, GRASIL (Silva et al. 1998). {\it We decided to investigate dust in the LMC because of the insights this galaxy can provide into the connections between star formation and the structure and dynamics of the ISM}. It is one of the nearest disc galaxies but is external to our own Milky Way galaxy, with a separation of 50 -- 55 kpc (Feast 1991; Alves et al. 2004), and provides us with a convenient laboratory for studying the environment of galaxies as well as the ISM and star formation feedback. The LMC is also oriented almost face-on, at 22$\pm6$$^{\circ}$ inclination from HI ATCA observations (Kim et al. 1998), and it has little foreground and internal extinction, which allows us to map the gas and dust of the ISM, making it possible to analyze stellar components and the ISM without any confusion. For these reasons, the LMC has been intensively studied at many wavelengths over the last few decades (Bothun \& Thompson 1988; Klein et al. 1989; Kennicutt et al. 1995; Oey \& Massey 1995; Haberl \& Pietsch 1999; Mizuno et al. 2001; Kim et al. 2003; Staveley-Smith et al. 2003; Zaritsky et al. 2004; Cioni et al. 2011). It has a number of features of interest, including active star formation in giant molecular clouds (GMCs) and supergiant shells (SGSs). The LMC also provides an excellent opportunity to study the effects of UV radiation from stars on their environment in the multi-phase ISM. The LMC is in a low metallicity environment with a small fraction of heavy elements, $Z=0.25$ $Z_\odot$ (Dufour 1984), $Z=0.3$--0.5 $Z_\odot$ (Westerlund 1997), and therefore contains fewer grain particles than the Milky Way. The total average star formation rate of the LMC, is relatively low as 0.26 $M_{\odot} yr^{-1}$ (Kennicutt et al. 1995), and is lower than that of the Milky Way as 0.68 -- 1.45 $M_{\odot}yr^{-1}$ (Robitaille \& Whitney 2010). This allows us to link studies of these interstellar properties to those of the early evolution of high-redshift and metal-poor galaxies.

In the present study, we explain the evolution of dust grains and how this process differs from that of the Milky Way, through analysis of dust models using different treatments of grain size distribution. We use submillimeter observations from the HSO and far-infrared observations from the SST to construct integrated SEDs for the LMC. We examine the effects of dust grain size distribution on the resultant SEDs. The remainder of this paper is organized as follows. In Section \S2, we briefly describe the characteristics of the data. In Section \S3, we describe modeling procedures together with the results of our calculation. In Section \S4, we discuss the results. A summary of the present study is provided in Section \S5.

\section{Data}

The Spitzer Space Telescope Legacy program "Surveying the Agents of Galaxy Evolution" (SAGE; Meixner et al. 2006) used the Infrared Array Camera (IRAC 3.6, 4.5, 5.8, 8.0 $\mu$m; Fazio et al. 2004) and the Multi-band Imaging Photometer for Spitzer (MIPS 24, 70, 160 $\mu$m; Rieke et al. 2004) to provide a uniform image of the LMC. The MIPS survey of the LMC was conducted as part of the SAGE Legacy program (Meixner et al. 2006). MIPS provides imaging capabilities at 24, 60, and 160 $\mu$m with bandwidths of 5, 19, and 35 $\mu$m, respectively. MIPS has two pickoff mirrors, which pass on the light to field mirrors at the back of the instrument for the 70 and 160 $\mu$m optical trains (Rieke et al. 2004).
The HERschel inventory of The Agents of Galaxy Evolution (HERITAGE; Meixner et al. 2010) also generated the SPIRE survey of the LMC at 250, 350, and 500 $\mu$m using the HSO (Pilbratt et al. 2010). We extracted an area within a 3.7 kpc circle and its center was on 79.284 deg. in R.A., --68.668 deg. in DEC.

Flux measurements for all infrared data were generated by convolving beam size to 38$''$ with FWHM. We convolved the Spitzer images by Gaussians using the SMOOTH task in MIRIAD (Sault et al. 1995), and processed the SPIRE (Griffin et al. 2010) images by using custom-made kernels (Gordon et al. 2008). On the other hand, Infrared Astronomical Satellite (IRAS) survey data provide a peak flux density at 100 $\mu$m, and data processing procedures are described in the IRAS explanatory supplement. Background levels of the IRAS observations were determined as 0.24 Jy/pixel for the 60 $\mu$m map and 0.57 Jy/pixel for the 100 $\mu$m map using the IRAS explanatory supplement. We also considered the effects of dust emission at millimeter and submillimeter wavelengths. Wilkinson Microwave Anisotropy Probe (WMAP) data observed at 23, 41, and 94 GHz from the WMAP 5-year release (Bennett et al. 2003) were converted to a 0.88 degree beam size in the $K$ band. Flux densities were measured for the observed extent of the LMC and the largest aperture size (0.88 degree beam) of the observed bands was used for all the bands. The background level was determined from the regions outside the aperture which was used to perform photometry at the given bands. The mean value was subtracted and the standard deviation is given as the photometric error in the fourth column of Table 1.

\section{Results}

Integrated SEDs over the multi-wavelengths from UV to far-infrared/submillimeter wavelengths are important for examining the physical properties of galaxies. These data
allow researchers to extract stellar parameters including star formation rates, and provide considerable information about the composition and abundance of interstellar dust. In this section, we investigate the integrated SEDs using observations made primarily by the HSO and SST, as well as other complementary datasets. Over the past few decades, various methods of calculating SEDs have been reported. The latest SEDs for the LMC were introduced by Israel et al. (2010), Kim et al. (2010), and Meixner et al. (2010). Galliano et al. (2011) probed non-standard dust properties and extended submillimeter excess using the HSO observations. The effects of dust on infrared emission can also depend on the morphological type of the galaxy (Dale et al. 2012). Skibba et al. (2012) examined the dust properties with stellar distribution of LMC using the resolved SEDs and found that the ratio of dust to stellar luminosity varies depending on the interstellar medium environment. In this section, we present the results of radiative transfer calculations using the spectrophotometric galaxy model, GRASIL developed by Silva et al. (1998), with modified grain size distributions for the dust in the LMC and resultant integrated SEDs.

\subsection{Grain Size Distribution}

The interstellar environment contains various types of interstellar dust with a range of different grain properties (Greenberg 1968). Distinct differences in grain size and relative abundance between the Milky Way and the Magellanic Clouds can be expected. In general, interstellar dust is formed from carbonaceous grains, called graphite or polycyclic aromatic hydrocarbons (PAHs), of the smallest grains and silicate grains. These grains of dust result in the absorption or scattering of incoming photons from background objects. The extinction of the interstellar radiation field in the LMC exhibits a spectral gap in the extinction curve for the Milky Way and the Small Magellanic Cloud (SMC) (Clayton et al. 2000; Weingartner \& Draine 2001). The galaxy exhibits an extinction of background radiation that is approximately inversely proportional to the wavelength. The distinctive absorption feature at 2175 \AA~ is thought to depend on the graphite component (Draine \& Lee 1984; Li \& Draine 2001). The extinction curve for the LMC generally exhibits a similar form at infrared and visible wavelengths although the bump at 2175 \AA~ is somewhat weaker, and the extinction increases sharply at far-UV wavelengths (Clayton \& Martin 1985). This indicates that graphite particles are on average slightly smaller and less abundant in the LMC than in the Milky Way. The typical value of the ratio of visual extinction to reddening, $R_v \equiv A(V)/E(B-V)$, is 3.1 -- 3.5 in the diffuse ISM and approximately 5 in the denser regions where grain sizes are larger due to accretion of dust from interstellar gas. The mean extinction curve in the LMC has been reported using $R_v=2.6$ (Weingartner \& Draine 2001), which is consistent with the diffuse interstellar environment of the LMC. The classical model of interstellar dust size distribution is based on the observed extinction of starlight along the diffuse line of sight. Mathis, Rumpl, \& Nordsieck (1977) suggested that the radiative effects of interstellar dust are dependent on the grain size distribution and the silicate and graphite composition. A typical standard distribution law is:

\begin{equation}
Dn_{gr}(a)=C n_H a^{-3.5} da, \, a_{min} < a < a_{max},
\end{equation}

where $n_{gr}$ is the number density of grains with size $\le a$, $n_H$ is the number density of H nuclei, $C=10^{-25.23}\, {\rm cm^{2.5}}$ for graphite and $10^{-25.11}\, \rm cm^{2.5}$ for silicate, and $a$ is the grain radius, where $50\, {\rm \AA} < a < 0.25\,\mu$m in the Milky Way. Here, we used the data reported by Piovan et al. (2006a) to calculate the extinction curve for the LMC using the modified Mathis, Rumpl, \& Nordsieck (MRN) model (1977). We adopted the functional form for the dust size distribution in the LMC from Weingartner \& Draine (2001). They calculated grain size distributions, including PAHs, consistent with the observed extinction for different values of $R_v$ in the local Milky Way and for regions in the LMC and SMC, by considering the different values of the total C abundance per H nucleus, $b_C$, in log normal components.

Piovan et al. (2006a) fitted the extinction curves for the Milky Way, LMC, and SMC by adopting the extinction curves from Weingartner \& Draine (2001) and minimising the $\chi^2$ error function. They also modified the power law of the MRN model by splitting the distribution law of the $i$-th component into several intervals. The dimensionless scattering and absorption coefficients, the ratio of $\sigma$ to $\pi$$a^2$, where $a$ is the dimension of the grain, including PAHs, silicate, and graphite grains were taken from Draine \& Lee (1984), Laor \& Draine (1993), Draine \& Li (2001), and Li \& Draine (2001) (and Piovan et al. 2006a).

The LMC is known to have low metallicity extragalactic environments. Therefore, according to Clayton et al. (2000), the dust grain size distributions for graphite and silicate may be reproduced for the extinction of the LMC along the line of sight. They concluded that the emission was best reproduced when the population of very small grains was the sum of two log normal size distributions as described further in Weingartner \& Draine (2001). According to these authors, the structure of the size distribution, $D(a)$, for the very small carbonaceous grains had only mild effects on extinction for wavelengths in the range of interest.

We calculated the size distributions for silicate and graphite grains in the LMC with $a_{0,1}$=3.5 \AA, $a_{0,2}=30$ \AA, $\sigma$=0.4, in the two log normal components ($b_{C,1}$=0.75$b_C$, $b_{C,2}$=0.25$b_C$) with the modified MRN model (Weingartner \& Draine 2001). The results are shown in Fig. 1 and summarized in Table 1. Diffuse ISM and dust are not yet understood. For example, there was a report on the non-detection of 10 $\mu$m silicate feature in the emission from diffuse clouds (Onaka et al. 1996). Li \& Draine (2001) proposed that emission was best reproduced if the very small grain population was the sum of two log-normal size distributions (Weingartner \& Draine 2001):

\begin{eqnarray}
\lefteqn{\frac{1}{n_H}(\frac{dn_{gr}}{da})_{vsg}\equiv D(a)}\nonumber\\
&&\qquad\qquad\qquad=\sum_{i=1}^{2}\frac{B_i}{a}exp\bigg\{-\frac{1}{2}\bigg[\frac{ln(a/a_{0,i})}{\sigma}\bigg]^2\bigg\},\nonumber\\
&&\qquad\qquad\qquad\qquad\qquad\qquad\qquad\qquad a > 3.5\AA,
\end{eqnarray}

\begin{eqnarray}
\lefteqn{B_{i}=\frac{3}{(2\pi)^{3/2}}\frac{exp(-4.5\sigma ^2)}{\rho a_{0,i}^3\sigma}}\nonumber\\
&&\qquad\times\frac{b_{C,i}m_C}{1+erf[3\sigma/\sqrt{2}+ln(a_{0,i}/3.5\AA/\sigma\sqrt{2})]},
\end{eqnarray}

where m$_C$ is the mass of a C atom, $\rho$=2.24$g$ $cm^{-3}$ is the density of graphite, b$_{C,1}$=0.75$b_C$, $b_{C,2}$=0.25$b_C$, $b_C$ is the total C abundance (per H nucleus) in the log-normal populations, a$_{0,1}$=3.5$\AA$, a$_{0,2}$=30$\AA$, and $\sigma$=0.4. We adopt the following form (Weingartner \& Draine 2001) for carbonaceous dust:

\begin{eqnarray}
\lefteqn{\frac{1}{n_H}\frac{dn_{gr}}{da}=D(a)+\frac{C_g}{a}(\frac{a}{a_{t,g}})^{\alpha_g}F(a;\beta_g,a_{t,g})}\nonumber\\
&&\qquad\times\left\{ \begin{array}{cc}
1 , & \\ {3.5\AA < a < a_{t,g}}\\
exp{-[((a-a_{t,g})/a_{c,g})]^3}, & \\ {a>a_{t,g}}
\end{array} \right\}
\end{eqnarray}
and
\begin{eqnarray}
\lefteqn{\frac{1}{n_H}\frac{dn_{gr}}{da}=\frac{C_g}{a}(\frac{a}{a_{t,s}})^{\alpha_g}F(a;\beta_s,a_{t,s})}\nonumber\\
&&\qquad\times\left\{ \begin{array}{cc}
1 , & \\ {3.5\AA < a < a_{t,s}}\\
exp{-[((a-a_{t,s})/a_{c,s})]^3}, & \\ {a>a_{t,s}}
\end{array} \right\}
\end{eqnarray}
for silicate dust.
Curvature can be provided by the following term (Weingartner \& Draine 2001):

\begin{eqnarray}
F(a;\beta,a_t)\equiv\left\{ \begin{array}{cc}
1+\beta a/a_t, & {\beta \ge 0}\\
(1-\beta a/a_t)^{-1}, & {\beta < 0}
\end{array} \right\}
\end{eqnarray}

For comparison, Fig. 1 also shows the size distributions for the Milky Way using the revised Draine \& Lee (1984) model adopted by Silva et al. (1998). The size of dust grains ranges from $a_{min}$ to $a_{max}$ and the exponents of $\beta_1$ and $\beta_2$ are free parameters with the values shown in Table 1. $b_C$ denotes the total $C$ abundance per $H$ in log normal components. The grain size distribution follows a power law but has a break that is introduced below $a_b$ where a larger exponent is used. The results are presented as Cases A and B (Weingartner \& Draine 2001) in Table 2 and Figs. 1 \& 2, which correspond to the different power law indices that we have used for the LMC. For silicate grains, varying the power law indices did not affect size distribution, whereas distinct variations appeared for graphite grains. The grain size distribution in Case B is in good agreement with the results of Piovan et al. (2006a). The size of the dust grains used in Piovan et al. (2006a) ranged from 1 nm to 300 nm for silicate grains and from 2 nm to 1 $\mu$m for graphite grains. The exponent of the best fit is -3.47 (Table 2 and Figs. 2 \& 3).

\subsection{Dust Properties}

Carbonaceous grains are diffused throughout the ISM due to outflows of the shells of late-type stars such as asymptotic giant branch (AGB) stars (Wood et al. 1983). Carbon stars in the LMC mainly consist of M-type stars, which have a low mass loss rate and low luminosity. As the formation efficiency of dust grains decreases, the graphite grain population in the interstellar environment decreases. This is also supported by the fact that the strength of the 2175 \AA~~bump in the extinction curve, which is caused by the graphite component of interstellar dust is reduced by a factor of 1.6, compared with that of the Milky Way.
Radiation in the near-infrared to far-infrared range is strongly affected by the size distribution of dust grains in the diffuse ISM (Draine \& Lee 1984; Kim et al. 1994; Li \& Draine 2001a). The effects of dust on infrared emission depend on chemical and physical composition, the star formation rate, the dust-to-gas mass ratio, and metallicity. Grains are also heated by starlight. While small grains composed of carbonaceous particles may explain the continuum emission in the mid-infrared and the 2175${\rm \AA}$ UV bump in the extinction curve, silicate materials can also have larger grains with diameters in the range of $\gtrsim$ 0.1 -- 2.0 $\mu$m according to the emission at far-infrared wavelengths (Papoular et al. 1996; Menella et al. 1998; Schnaiter et al. 1998; Gordon et al. 2003; Steglich et al. 2010). The infrared spectra of a wide variety of sources are dominated by strong emission line features at 3.3, 6.2, 7.7, 8.6, 11.3, and 12.7 $\mu$m. These are generally attributed to PAHs or PAH-related molecules (Leger \& Puget 1984; Allamandola et al. 1985; Sellgren et al. 1988; Schutte et al. 1993; Tielens \& Snow 1995; Allain et al. 1996a,b; Boulanger et al. 1998; Bakes et al. 2001; Hony et al. 2001; Verstraete et al. 2001; Pech et al. 2002; Peeters et al. 2002; Hudgins \& Allamandola 2004; Madden et al. 2006; Galliano et al. 2008). The profiles, relative strengths, and peak positions of these features are determined by local conditions and processes. Amorphous silicates reside principally in a galaxy's interstellar clouds. If they strongly absorb radiation from the surrounding interstellar environment, they can transform into crystalline silicates with a solid arrangement. Crystalline silicates (Kemper, Vriend, \& Tielens 2004; Speck, Whittington, \& Tartar 2008) that appear in AGB stars (Matsuura et al. 2009; Srinivasan et al. 2009) are easily found in the discs of young stellar objects that are relatively small. The outflows of evolved stars and planetary nebulae (PNe) also provide sources of crystalline silicates (Waters et al. 1996; Molster et al. 2000; Spoon et al. 2006). The crystallization and grain growth is done at early stages of disc formation and evolution (Williams \& Cieza 2011). Because the star formation rate (SFR) of the LMC is known to be about 0.26 -- 0.29  $M_\odot$/yr  which seems to be lower than the average value of the SFR of the Milky Way, it is expected to have a lower abundance of crystalline silicates in the LMC. This means not only that silicate abundance is reduced, but also that the young stellar object (YSO) (Whitney et al. 2008; Gruendl \& Chu 2009) formation rate, which is a source of forming crystalline silicates (Natta et al. 2007; Voshchinnikov \& Henning 2008), is slower.

\subsection{Spectral Energy Distributions}

Approximately 30\% of all light radiated by stars in the local universe has been reprocessed by dust (Soifer et al. 1991). Measurements of far-infrared/submillimeter backgrounds suggest that 50\% of the light radiated by stars has been reprocessed over the entire history of the universe (Hauser \& Dwek 2001). An appropriate treatment of dust reprocessing in galaxies is therefore essential to determine the physical quantities related to star formation histories based on observed data, and to test theoretical galaxy formation models against observations. The effects of dust on the infrared emission may vary depending on the morphological type of the galaxy. Submillimeter excess is presented in dwarf irregular galaxies and low metallicity systems (Galametz et al. 2011; Galliano et al. 2011; Dale et al. 2012). The presence of very cold dust, submillimeter emissivity depending on the temperature, and/or combination of spinning and thermal dust emission are suggested as the origin of the submillimeter excess, but the cause of submm excess still remains in question. In this section, we describe integrated SEDs using a modified grain size distribution for the LMC and show the results of radiative transfer calculations using the spectrophotometric galaxy model, GRASIL (Silva et al. 1998). We adopted a modified grain size distribution for the dust model for the LMC that was described in the previous section. We present the integrated SEDs and the results of radiative transfer calculations in Figs. 2 -- 6.

Integrated SEDs over infrared to submillimeter wavelength ranges can elucidate the physical properties and identification of galaxies, including stellar parameters such as SFRs, and parameters describing the composition, abundance, and physical structure of the ISM. Over the past few years, various models have been proposed to describe SEDs (Silva et al. 1998; Granato et al. 2000; Piovan et al. 2006a,b). Galliano et al. (2011) and Skibba et al. (2012) presented the most recent model for describing SEDs for the LMC. Israel et al. (2010) probed submillimeter and centimeter excess of Magellanic Clouds with integrated full SED covering radio to ultraviolet. A weak mid-IR excess is shown in the LMC, and Israel et al. (2010) attribute this to the lack of PAHs by the low metallicity, strong radiation field, strong shocks, and destruction of PAHs.

Table 2 lists the parameters from the modified MRN model using the best fitting SEDs from the observational data for the LMC and the Milky Way. Cases A and B incorporate the power law indices of the size distributions that we have taken for the LMC; these are summarized in Table 2, where we adopted the grain size distributions including PAHs that were consistent with the observed extinction for different values of $R_v$ in the local Milky Way and for the regions in the LMC and SMC, taking into account $b_C$ (Table 2). Weingartner \& Draine (2001) calculated grain size distributions including PAHs that were consistent with the observed extinction for different values of $R_v$ for the regions in the LMC. Piovan et al. (2006a) were able to fit the extinction curves for the Milky Way, LMC, and SMC by minimizing the $\chi^2$ error function. Here, we used the results of Piovan et al. (2006a) to calculate the LMC extinction curve using the modified MRN model. We estimated parameters for the LMC using the grain size distribution of Weingartner \& Draine (2001) and a radiative transfer calculation. For silicate grains, varying the power law indices resulted in very small changes in the model SED within the observational uncertainties. Case B was generally in good agreement with the results of Piovan et al. (2006a) and Silva et al. (1998) for graphite grains, mainly due to splitting the size components into several levels.

GRASIL is a fit model that computes the sizes of dust grains semi-analytically (Silva et al. 1998; Granato et al. 2000). The effects of a dusty ISM on the environment are also considered in the radiative transfer calculation. For the LMC, we used a standard radiative transfer model with the dust models that we derived in the previous section. The results of the calculation are given in Fig. 2. The contribution of each component of the ISM was also measured and the results are shown in Fig. 2. Here, we describe the model and the input parameters we used to fit it. GRASIL calculates the radiative transfer of the starlight, heating of dust grains, and the emission from these grains with a specific grain model, and grain temperatures for the geometrical distribution of the stars and dust (Silva et al. 1998; Granato et al. 2000).

Geometry: In the present calculation we considered a disc system with a distribution of stars, gas, and dust. The modelling is performed in spherical coordinates and we assume azimuthal symmetry with respect to the equatorial plane. The radial scale-lengths of the exponential disc in the galaxy was estimated as $log_{10}(R_d/kpc)\sim-0.2\,M_B-3.45$ with $R_d\sim1.8$ kpc and $M_B=-18.57$ (de Vaucouleurs \& Freeman 1972). The scale-heights of the disc were approximated to be about 180 pc. The spatial distribution of the three components, i.e., stars, molecular clouds, and the diffuse ISM were fixed by the radial scale lengths of $R^*_d \approx R^{MC}_d \approx R^{ISM}_d$ and followed a simple assumption that the vertical spatial scales for the three components were almost the same in order to reduce the number of scale parameters, firstly. Then, we also modelled the SEDs by adopting $z_*\approx550$ pc and $z_d\approx180$ pc (Kim et al. 1999; references therein). We found that the variance of the SED fit and the observed data set was reduced after we gave different scale-heights for the dust (and gas) from the vertical scale-height of stars in the disc. These results are shown in Figs. 3 and 5. The inclination of the disc was taken as about 30 degree in order to include both gas (dust) and stellar discs.

Synthesis of the Starlight Spectrum: The luminosities of different stellar components in the galactic disc and young stars still in the clouds were calculated using the population synthesis model with convective overshooting from the Padova library (Bertelli et al. 1994) together with the isochrones from Tantalo et al. (1998)(and Piovan et al. 2006a). The initial masses of the evolutionary tracks ranged from 0.15 M$_\odot$ to 120 M$_\odot$, corresponding to ages ranging from a few Myrs to several Gyrs (Piovan et al. 2006a). We adopted a Salpeter initial mass function (IMF) of $\psi(m)\propto M^{-x}$ ($x=2.35$ for 0.15 $M_\odot < m < 120 M_\odot$).

Optical Depth: The optical depth of a molecular cloud especially affects the SEDs of young stars in the library (Granato et al. 2000). As noted above, the dust in the GRASIL originates from two components which are dense molecular clouds and diffuse cirrus clouds in the disc. If the optical depth is high, the energy becomes shifted towards longer wavelengths (Piovan et al. 2006a). Optical depth increases with mass but decreases with size, because the optical depth is proportional to $m_{MC}/r^2_{MC}$ where $m_{MC}$ is the mass of an individual molecular cloud in a galaxy and $r_{MC}$ is the radius of each cloud in the disc (Silva et al. 1998). Thus, the size of molecular cloud and the mass of each cloud governs the optical depth. We can choose $\tau_V\approx5$ (for the SMC), $\tau_V\approx10$ (for the LMC), or $\tau_V\approx35$ (for the Milky Way) for the library of stars following Piovan et al (2006a). This is similar to the optical depth at 4--5 $\mu$m which is about 0.3--0.5 and is coincident with the value we obtain by fitting the observed SEDs under the assumption that $m_{MC}$ is approximately 10$^6M_\odot$ (Yamaguchi et al. 2001a,b) and the average size of the CO cloud is approximately 40 pc (Fukui et al. 2009).

Escape Time Scale: A time scale $t_{esc}$, controls when young stars usually escape from their parental clouds in the galactic disc (Silva et al. 1998). The present model allows $t_{esc}$ to take different values in normal discs and in bursts. In normal galaxies, star formation takes place in clouds throughout the disc, and young stars are assumed to be distributed throughout the disc after they escape from their parental clouds (Granato et al. 2000). A value of 2 Myrs can be used for irregular galaxies based on the experimental tests (Piovan et al. 2006a). Silva et al. (1998) and Granato et al. (2000) reported that relatively larger values for $t_{esc}$ between 20 and 60 Myrs are needed to fit the SEDs of typical starburst galaxies. High values of $t_{esc}$ mean that young stars are hidden longer by the parental clouds and much of the light from young stars emitted at UV and optical wavelengths is shifted to the far infrared by absorption and re-emission within the dust cloud. A large fraction of the infrared light emitted by a galaxy could be due to young stars that are still embedded in the molecular clouds (Silva et al. 1998; Granato et al. 2000; Piovan et al. 2006a).

Dust Abundance: The dust-to-gas ratio was assumed to be proportional to the gas metallicity and defined as $\delta\approx M_d/M_H$ where $M_d$ is the total dust mass and $M_H$ is the total gas mass (Silva et al. 1998; Granato et al. 2000). We used $\delta\sim1/300$ for the LMC, which is about a median value across the literature reporting dust-to-gas ratios for the LMC (Pei 1992; Bernard et al. 2008; Roman-Duval et al. 2010). The dust-to-gas mass ratio ranges between $\sim$ 0.02 and 0.0002 in the gas-to-dust ratio map of Galliano et al. (2011), while Skibba et al. (2012) showed resolved dust properties of the Magellanic Clouds from the resolved SEDs and gave the total gas-to-dust mass ratio of $340 \pm 40$. The relation $\delta\approx\delta_\odot$ ($Z/Z_\odot$), incorporating the effects of metallicity was adopted to evaluate the amount of dust in the galaxy models (Silva et al. 1998; Granato et al. 2000; Piovan et al. 2006a). Based on the previous studies and reports on this matter, $\delta$ varies from 0.01 to 0.002 for the Milky Way and other galaxies of the Local Group. This relation implies that relatively metal poor galaxies differ in terms of both the abundances of heavy elements and the relative proportion of the dust grains and diverse patterns in chemical compositions of the dust grains. We tested the effects of dust abundance on the SEDs and the results were given in Fig. 5. The dust-to-gas ratios cause metallicity differences and hence affect the pattern of SEDs. To probe this further, the SEDs were fitted for the dust amount of $\delta\approx\delta_\odot$($Z/Z_\odot$). Fig. 5 presents the results: dust-to-gas ratio strongly affects SEDs, particularly at wavelengths near to 100 $\mu$m, and higher dust-to-gas ratios result in lower fluxes in the mid- and far-infrared wavelength range.

Dust Model for the ISM: The effects of dust on the radiative transfer calculation depend on the physical and chemical properties of dust grains (Silva et al. 1998). The main constituents of dust in the disc are molecular clouds and dust surrounding YSOs, dust in the diffuse ISM, and stellar outflows (Dorschner \& Henning 1995). Predictions based on Mie theory usually work better for longer wavelengths than shorter wavelengths typically from infrared to the optical wavelengths. According to Mie theory, the extinction curve rises sharply as the wavelength decreases. We were educated that the existence of the bump in the extinction curve provided us with information about the composition of interstellar dust and its size distribution. Here, we review the characteristics of dust model that we briefly noted above. In general, interstellar dust is formed from carbonaceous grains called graphite or polycyclic aromatic hydrocarbons (PAHs) of the smallest grains and silicate grains, which give rise to absorption or scattering of incoming photons. The extinction of the interstellar radiation field in the LMC exhibits a gap in the extinction curve for the Milky Way and the SMC. The Milky Way and the SMC galaxies exhibit an extinction that is approximately inversely proportional to the wavelength, following Mie theory. Understanding extinction is important because classical models of interstellar dust size distribution are based on the observed extinction of starlight along the diffuse line of sight. The distinctive absorption feature at 2175 \AA~ depends on the graphite composition of the ISM (Draine \& Lee 1984; Li \& Draine 2001). The extinction in the LMC shows a similar pattern at infrared and visible wavelengths, although it has a rather weak bump at 2175 \AA~ and increases sharply at far-UV wavelength in the LMC (Clayton \& Martin 1985). This indicates that graphite particles are on average slightly smaller and less abundant in the LMC than in the Milky Way (Clayton et al. 2000; Weingartner \& Draine 2001).

In the present study, we adopted the results of Piovan et al. (2006a) to calculate the LMC extinction curve with the modified Mathis, Rumpl, \& Nordsieck (MRN) model (1977). The LMC is known to have a lower metallicity extragalactic environment than that of our Milky Way (Russell \& Dopita 1992). The dust grain size distribution for graphite/silicate was reproduced for the extinction of the LMC along the line of sight. Clayton et al. (2000) concluded that the emission was best reproduced when the small grain population had the sum of two log-normal distributions. The structure of the size distribution, $D(a)$ for the very small carbonaceous grains only had mild effects on the extinction for the wavelengths of interest. We calculated the size distributions for silicate and graphite grains in the LMC by adopting the values for $a_{0,1}$=3.5 \AA, $a_{0,2}=30$ \AA, $\sigma$=0.4, in the two log normal components ($b_{C,1}$=0.75$b_C$, $b_{C,2}$=0.25$b_C$) with the modified MRN model (Weingartner \& Draine 2001; Piovan et al. 2006a) as noted above. The sizes of dust grains ranged from $a_{min}$ to $a_{max}$, and the exponents of $\beta_1$ and $\beta_2$ are free paramters. $b_C$ denotes the total $C$ abundance per $H$ in the log normal components. The carbon abundance, $b_C$ per H nucleus in two log-normal populations is also an important parameter, fixes the abundance of the element in the two log normal populations of very small grains as discussed by Weingartner \& Draine (2001). The PAH emission in the mid-IR approached higher flux levels for higher values of $b_C$. However, relatively low values of $b_C$ increased  the emission in the far-IR wavelength range. We derived normalization factor and index of $-3.47$ -- $-3.45$ with grain radii from 5 nm to 250 nm. The grain size distribution followed a power law but had a break to a steeper power law which was introduced below $a_b$. Case A and B correspond to the different power law indices, respectively, that we have taken for the LMC. The grain size distribution in Case B shows a generally close agreement with the results of Piovan et al. (2006a). The ionization of PAHs also affects the SEDs (Silva et al. 1998). The ionization state of PAHs reportedly changes emission profiles in the mid-IR (Weingartner \& Draine 2001; Piovan et al. 2006a).

\section{Discussion}

Here, we summarize the effects of the dust size distribution on the SEDs. The effects of grain size distribution on the SEDs are shown in Fig. 3, where the different power law indices are noted. The contribution of each component of the ISM is shown in Figs. 3 and 4. The parameters with the best fit using the GRASIL model were similar to those used by Piovan et al. (2006a). Table 2 lists the results of two separate SED fitting processes, achieved by varying the abundance of the size distributions of the dust grains in the medium, are listed in Table 2. The effects of these parameter changes in the size distributions related to Case A and Case B on the SEDs were most apparent at far-infrared wavelengths. Changes in the size distributions of silicate grains did not significantly affect the resultant SEDs, but similar changes in the graphite grain distributions did. The SEDs that fitted the observed data using the different size distribution reveal that power law indices for carbonaceous dust grains differed significantly, especially in the far-infrared wavelength range (see Fig. 3).

We were able to probe a parameter set in chemical and geometric space. Based on these parameters, which have been detailed above, we were able to confirm that the grain properties and dust size distributions used in this study are plausible for modeling the SEDs. The dust content of the LMC also affects the SEDs as we described above. The dust-to-gas ratios result in different metallicity and hence affect the pattern of the SEDs. Here, the dust-to-gas ratio is defined as $\delta\approx M_d/M_H\approx\Sigma_{i}m_{i}/M_H$ where $m_i$ is the mass of the $i$-th grain type. The mass $m_i$ can be obtained by integrating over the grain size distribution which is given as a function of the abundance coefficients. Depending on the age of the galaxy, the gas content is known and the amount of dust in the ISM can be given as a function of $\delta$, which is reportedly in the range of 0.002 -- 0.01 for the Milky Way and other galaxies of the Local Group. A simple relation incorporating the effects of metallicity, $\delta\approx\delta_\odot(Z/Z_\odot)$, has been used to evaluate the amount of dust in the galaxies (Silva et al. 1998; Granato et al. 2000; Piovan et al. 2006a). Differences in metallicity imply differences in the abundances of heavy elements and differences in the relative proportions and chemical compositions of dust grains. As shown in Fig. 5, the dust-to-gas ratio specially affects the SEDs in the mid- and far-IR range. When a higher dust-to-gas ratio is chosen, this results in a smaller flux at wavelength especially near 100 $\mu$m. Related to the metallicity of a galaxy's environment, knowledge on the relative proportions of the various components of dust grains is needed to understand the evolution of dusty environment and complete understanding of dust yields (Zubko, Dwek, \& Arendt 2004) is also critical (Galliano et al. 2011). There are some studies on the observed correlation between the strength of PAH features and metallicity of galaxies (Madden et al. 2006) and there were also several efforts on explaining these relation between PAH abundance and metallicity using chemical evolution models, as reported in Galliano et al. (2008).

Zubko et al. (2004) presented a comprehensive dust model including PAHs, silicates, graphites, amorphous carbon, and composite particles and solved for the optimal grain size distribution of each dust component for the Milky Way. However, they concluded that their modeling results could not provide a unique dust model that fits all of these, constraints on elemental abundances, infrared emission, and extinction. They presented 15 different cases for their dust models. Galliano et al. (2011) chose one of Zubko et al. (2004) dust model which called the bare grain model with solar abundance constraints, BARE-GR-S model in their analysis of the SEDs of the LMC. This model includes a higher small grain contribution than that of Draine \& Li (2007). Galliano et al. (2011) considered this model because the fit of the extinction curves of the LMC indicated a larger fraction of smaller grains based on the studies by Weingartner \& Draine (2001). But in their studies the model was modified to lower the abundance of non-PAH small grains where their sizes for both carbon and silicate grains were less than 10 nm by a factor of 2 in order to achieve the best model fits to the infrared emissions at MIPS bands, especially at 24 $\mu$m. Galliano et al. (2011) reported that the use of integrated SED causes the underestimation of dust mass by a factor of
two and this might be due to the dilution of cold massive
regions in the case of relatively low resolution maps.

Table 2 and Figs. 1 -- 3, we present the size distributions for silicate and graphite grains in the LMC by Weingartner \& Draine (2001) and the modified MRN models used in Piovan et al. (2006a), which we also used for the SEDs of the LMC in the present
study. Cases A and B correspond to the different power law indices,
respectively, which we have also taken for the LMC. Fig. 4 presents the SEDs of the LMC with different power law indices for the size distribution of carbonaceous dust
grains. This effect mainly appears at wavelength ranges
from mid- to far-infrared wavelengths, and is especially
distinct in the far-infrared wavelength range. Changing
the power law indices for the silicate grains had no
significant effects in terms of parameter changes on the
SEDs. The dust content of the LMC also affects the SEDs
as noted above. The dust-to-gas ratios, which are given
as the metallicity, cause the variation in the model fits
shown in Fig. 5. The difference mainly occurs at
wavelengths near 100 $\mu$m and longer than 1 mm, as noted
above. The results of this study suggest that the
dust-to-gas ratio we adopted to model the SEDs, $\approx$
1/300 gives a reasonable fit to the observed SEDs and provide
a coincident result with the metallicity of the LMC (Russell \& Bessell 1989) for a given synthesis of starlight spectrum,
optical depth, escape time scale, and dust model. Fig. 6 presents the star formation efficiency (SFE) used to fit
the observed SEDs. First, we used a star formation
efficiency of $\nu\approx0.05$ Gyr$^{-1}$ (as
suggested by Piovan et al. 2006a), which was an input
parameter to model the SEDs. For the parameters described
above, the best model fit to the observed SEDs was obtained
with a star formation efficiency of 0.08 Gyr$^{-1}$, and
the resultant star formation rate with 0.29 $M_\odot$/yr,
which is close to the star formation rate of the LMC
(Kennicutt et al. 1995). Decreasing the star formation
rate and the Schmidt efficiency decreases the overall flux
densities in the SEDs, except in the mm wavelength range.
Thus, we conclude that we could probe a reasonable set of
parameter values in chemical and geometric space with the
dust size distribution of a modified MRN distribution with
the LMC extinction curve. We were also able to reproduce
the observed SEDs of the LMC.

\section{Summary}

We investigated the integrated SEDs from infrared to
submillimeter emissions in the LMC and described the results of radiative transfer calculations using the spectrophotometric galaxy model, GRASIL. We explained the data using dust models with differing treatments of grain size distributions because the LMC has undergone different processes from the Milky Way. We used submillimeter observations and far-IR observations from the HSO and mid-IR observations taken with the SST. Interstellar dust grains are formed from graphite and PAHs, as well as silicate grains, which cause extinction of incoming photons from the background objects. We calculated the LMC extinction curve using the modified MRN model (Piovan et al. 2006a) and grain size distributions by Weingartner \& Draine (2001). We adopted the same values as Li \& Draine (2001) and Weingartner \& Draine (2001) for $a_{0,1}$=0.35 nm, $a_{0,2}$=3 nm, $\sigma$=0.4, and the same relative populations in the two log-normal components ($b_{C,1}$=0.75$b_C$, $b_{C,2}$=0.25$b_C$). The best fit of the observed SEDs for the LMC indicate that the grain size distribution follows a relatively less steep power law than that of the Milky Way but has a break that is introduced below $a_b$, where a larger exponent is used. For silicate grains, varying the
power law indices caused minor changes in the model SEDs within the observational uncertainties, whereas changing graphite grain size distribution did significantly change
the pattern of SEDs. The model SEDs that fitted the
observational data using different size distributions suggest
that the power law indices for carbonaceous dust grains
differ significantly, especially in the far-infrared
wavelength range.

Based on the grain size distributions listed in Table 2, we
were able to probe an appropriate parameter set describing
the LMC in chemical and geometric space by modeling the
SEDs using the GRASIL (Silva et al. 1998), a spectrophotometric galaxy model, by performing radiative transfer calculations. In this calculation, we considered a disc system with a distribution of stars, gas, and dust, and modelled the SEDs by assuming that 1) the vertical scale heights of gas (dust) and stars are the same and
2) vertical scale heights of gas (dust) and stars are
different. For the case 2) we found that the variances of
the SEDs fit and the observed datasets were reduced. For a
given set of parameters, including synthesis of the starlight
spectrum, optical depth, escape time scale, and dust model
described in Section 3 and 4, the adopted dust-to-gas ratio for
modelling the SEDs provided a coincident result with the metallicity
of the LMC reported by Russell \& Bessell (1989). The dust-to-gas ratios
that are given as the metallicity caused the variation in the model fits,
as depicted in Fig. 5. Differences mainly appear at wavelengths near 100
$\mu$m based on the dust models used to fit the observed SEDs in GRASIL.
Overall, we were able to reproduce the observed SEDs of the LMC with a reasonable
set of parameter values in chemical and geometric space, together with the
dust size distribution of a modified MRN model with the LMC extinction curve.

This work was made possible in part by the use of data products
published in A\&A (2010). We thank the PIs of the projects and all the team members.
We thank an anonymous referee and editors for very helpful comments to improve
the manuscript in its original form. Herschel Space Observatory is operated by the European Space Agency with science instruments
provided by European-led Principal Investigator consortia, JPL, and NASA. This research was supported in part by Mid-career Researcher Program through the National Research Foundation of Korea (NRF) funded by the Ministry of Education, Science, and Technology 2011-0028001.

\nocite{*}
\bibliographystyle{spr-mp-nameyear-cnd}

\newpage

\begin{table}
\caption{Observed flux densities}
\begin{tabular}{@{}rrcr@{}}
\tableline
$\nu$ & $\lambda$ & Flux density &  \\
$10^2$GHz & [$\mu$m] & [kJy] & [$\triangle$kJy]\\
\tableline
832.75 & 3.6 & 2.40 & 0.12\\
666.20 & 4.5 & 1.64 & 0.08\\
516.88 & 5.8 & 3.66 & 0.18\\
374.74 & 8.0 & 6.39 & 0.32\\
124.91 & 24.0    & 8.10   & 0.16 \\
49.97  & 60.0    & 93.12  & 18.62\\
42.83  & 70.0    & 137.47 & 6.87\\
29.98  & 100.0   & 242.79 & 48.56\\
18.74  & 160.0   & 294.72 & 35.37\\
11.99  & 250.0   & 158.98 & 31.80\\
8.57   & 350.0   & 75.06  & 15.01\\
6.00   & 500.0   & 37.59  & 7.52\\
0.94   & 3200.0  & 0.35   & 0.11\\
0.41   & 7300.0  & 0.16   & 0.05\\
0.23   & 13000.0 & 0.18   & 0.05\\
\tableline
\end{tabular}
\end{table}

\begin{table}
\small
\caption{Parameter Values of Grain Size Distribution}
\begin{tabular*}{.48\textwidth}{@{}rccccc}
\tableline
& MW & MW & LMC & LMC & LMC \\
& DL & S98 & Piovan & Case A & Case B \\
\tableline
&&&Silicate&&\\
\tableline
logC&  -25.11&    -25.11&     -25.35&     -25.31&      -25.55 \\
a$_{min}\,{\rm [\AA]}$&  50&    -&      10&     50&     50 \\
a$_{max}\,{\rm [\AA]}$&  2500&  2500&   3000&      2500&       2500 \\
a$_{b}\,{\rm [\AA]}$&    -&     50&     -&      -&      - \\
$\beta_1$&  -3.5 &   -3.5 &   -3.47&  -3.45&   -3.5 \\
$\beta_2$&  -&  -&  -&  -&  -\\
\tableline
&&&Graphite&&\\
\tableline
logC&     -25.13&     -25.22&     -25.96&     -25.75&     -25.90 \\
a$_{min}\,{\rm [\AA]}$&     50&     8&      20&     50&     50 \\
a$_{max}\,{\rm [\AA]}$&     2500&       2500&       10000&      2500&       2500 \\
a$_{b}\,{\rm [\AA]}$&       -&      50&     200&        -&      - \\
$\beta_1$&  -3.5&   -3.5&   -3.5&   -3.45&   -3.5 \\
$\beta_2$&  -&  -4.0&   -4.0&   -&  - \\
\tableline
\end{tabular*}
\label{tab:GSD}
\end{table}

\begin{figure}
\vspace{0.5cm}
\hspace{-0.3cm}
\includegraphics[width=0.49\textwidth]{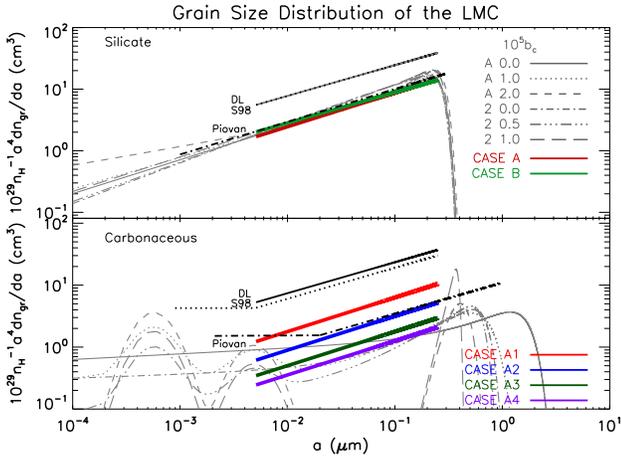}
\caption{Grain Size Distribution of the LMC. The upper diagram exhibits the extinction of the silicate grains and the lower diagram of the carbonaceous grains such as graphite. $10^{5}\,b_{C}$ is related to the total C abundance per H nucleus in log normal components. Diverse gray lines are for the LMC: "A" denotes the distribution for the LMC average extinction and the "2" denotes the distribution for the LMC2 area as given in Weingartner \& Draine (2001). Piovan (black dash-dot line), Case A (red solid line) and Case B (green solid line) correspond to the power law index of size distribution for the LMC, as given in Table 2. DL (black solid line) and S98 (black dotted line) correspond to the size distributions for the Milky Way.}
\end{figure}

\begin{figure}
\vspace{0.5cm}
\hspace{-0.3cm}
\includegraphics[width=0.49\textwidth]{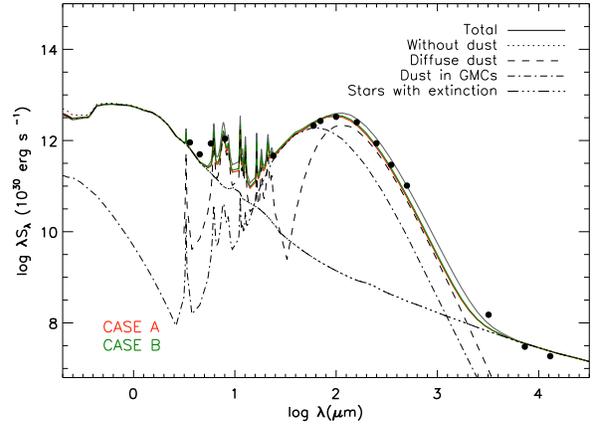}
\caption{The best-fit SED of the LMC using the radiative transfer calculation with GRASIL (Silva et al. 1998). The black filled circles are the observational data from the Spitzer Space Telescope (SST) and the Herschel Space Observatory (HSO). The lines represent contributions of the SEDs without dust (dotted), of diffuse dust (dashed), of dust in the GMCs (long dashed-dot), and of stars with extinction (long-dashed) to the total SEDs (solid). CASE A and CASE B indicate each size distribution noted in Table 2 and the results denoted as black solid line were derived using the size distibution, similar to that used in Piovan et al. (2006a). Here we set $z_*\approx z_d\approx180$ pc for the scaleheights of stars ($z_*$) and dust ($z_d$), respectively. We assume that the scale-height of the gas distribution in the disc is approximately similar to that of the dust distribution in the disc.}
\end{figure}

\begin{figure}
\vspace{0.5cm}
\hspace{-0.3cm}
\includegraphics[width=0.49\textwidth]{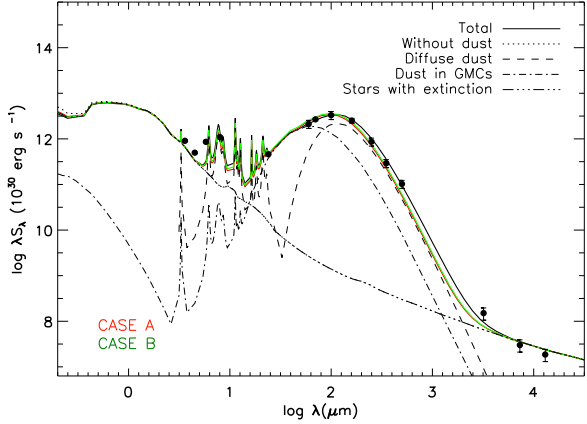}
\caption{The best-fit SED of the LMC with our observational data within uncertainties. The black filled circles are the observational data from the Spitzer Space Telescope (SST) and the Herschel Space Observatory (HSO).
The lines represent contributions of the SEDs without dust (dotted), of diffuse dust (dashed), of dust in the GMCs (long dashed-dot), and of stars with extinction (long-dashed) to the total SEDs (solid). We set $z_*\approx550$ pc and $z_d\approx180$ pc for the scale-heights of stars and dust (gas), respectively. We assume that the scale-height of the gas distribution in the disc is approximately similar to that of the dust distribution in the disc.}
\end{figure}

\begin{figure}
\hspace{-0.3cm}
\includegraphics[width=0.49\textwidth]{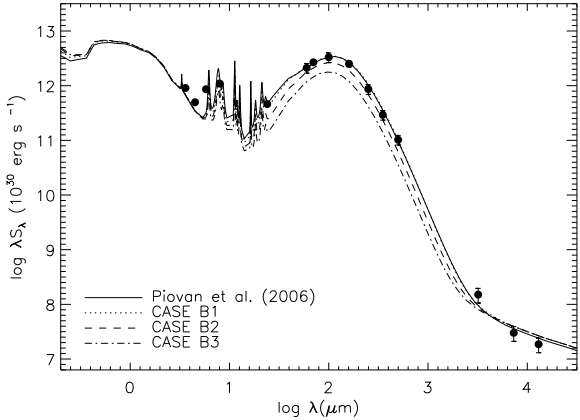}
\caption{The SEDs of the LMC with different power law indices
  of the size distribution of carbonaceous dust grains. Case B1
  corresponds to the parameters of Case B in Table 2. The
  power indices change from $3.5$ to $3.4$ by 0.5. For a comparison the best fit of SEDs are given as the dotted line. The difference in the SEDs is seen mainly in the Mid-infrared and Far-infrared range.
  }
  \end{figure}

\begin{figure}
\vspace{0.5cm}
\hspace{-0.3cm}
\includegraphics[width=0.49\textwidth]{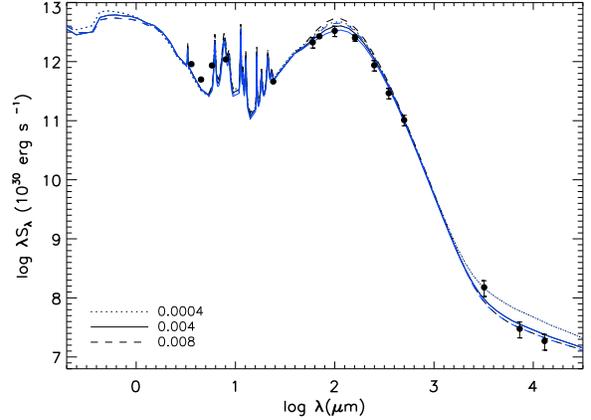}
\caption{The SEDs of
the LMC for the different metallicity. The metallicity effects on the SEDs are depicted as the dust amount, $\delta = \delta_\odot(Z/Z_{\odot})$. Difference of the effects on the SEDs is noted for the wavelength at near $\sim 100\,\mu{\rm m}$ as well as mid- to far-IR range. Blue lines indicate different disc scale-heights of dust and stars, $z_d\approx180$ pc and $z_*\approx550$ pc respectively.
}
\end{figure}

\begin{figure}
\vspace{1.0cm}
\hspace{-0.3cm}
\includegraphics[width=0.49\textwidth]{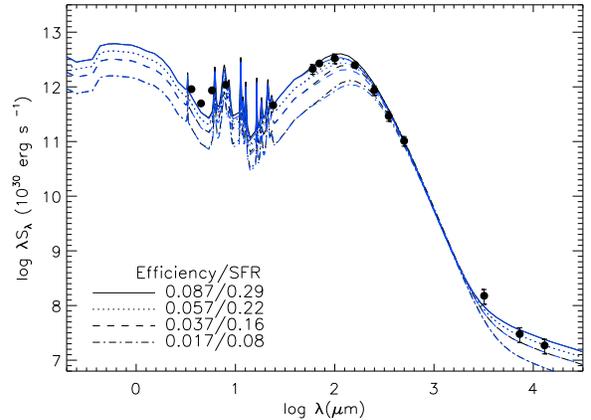}
\caption{The SEDs of the LMC for the different star formation rates (SFRs). The SFRs and the star formation efficiency of the Schmidt law are given by decreasing from 0.29 $M_\odot/yr$ for the best-fit SEDs (solid line) to 0.08 $Gyr^{-1}$ (dash-dotted line). The SFR seems to have no affection on the characteristic feature of the SEDs but shows shift in the SEDs via changing SFRs. Blue lines indicate different disc scale-heights of dust and stars, $z_d\approx180$ pc and $z_*\approx550$ pc respectively.
}
\end{figure}

\end{document}